\documentclass[letter,twocolumn]{emulateapj}
\usepackage{graphicx, txfonts}
\usepackage{url}
\usepackage{multirow}
\usepackage{booktabs}

\usepackage[usenames,dvipsnames]{color}

\newcommand{\DI}{\textrm{D}\,\textsc{i}}
\newcommand{\HI}{\textrm{H}\,\textsc{i}}
\newcommand{\HeI}{\textrm{He}\,\textsc{i}}

\newcommand{\HII}{\textrm{H}\,\textsc{ii}}
\newcommand{\ob}{$\Omega_{\rm B,0}$}
\newcommand{\obhh}{$\Omega_{\rm B,0}\,h^2$}
\newcommand{\neff}{$N_{\rm eff}$}
\newcommand{\heir}{$^{3}{\rm He}/^{4}{\rm He}$}
\newcommand{\yp}{Y$_{\rm P}$}

\shorttitle{\textsc{The Helium Isotope Ratio}}
\shortauthors{\textsc{Cooke}}

\begin{document}

\title{Big Bang Nucleosynthesis and the Helium Isotope Ratio}

\author{Ryan J. Cooke\altaffilmark{1,2,3}}

\altaffiltext{1}{Department of Astronomy and Astrophysics, University of California, 1156 High Street, Santa Cruz, CA 95064, USA}
\altaffiltext{2}{UCO/Lick Observatory, University of California, Santa Cruz, CA 95064, USA}
\altaffiltext{3}{Hubble Fellow;~~~~email: rcooke@ucolick.org}

\begin{abstract}
The conventional approach to search for departures from the standard model
of physics during Big Bang Nucleosynthesis involves a careful, and subtle measurement
of the mass fraction of baryons consisting of helium. Recent measurements of
this quantity tentatively support new physics beyond the standard model but,
historically, this method has suffered from hidden systematic uncertainties.
In this letter, I show that a combined measurement of the primordial deuterium
abundance and the primordial helium isotope ratio has the potential to provide a
complementary and reliable probe of new physics beyond the standard model.
Using the recent determination of the primordial deuterium abundance and assuming
that the measured pre-solar \heir\ meteoritic abundance reflects the primordial value,
a bound can be placed on the effective number of neutrino species,
\neff(BBN)~$=3.01^{+0.95}_{-0.76}$ (with 95 per cent confidence).
Although this value of \neff\ supports the standard model, it is
presently unclear if the pre-solar \heir\ ratio reflects the primordial value. New
astrophysical measurements of the helium isotope ratio in near-pristine environments,
together with updated calculations and experimental values of several important
nuclear reactions (some of which are already being attempted), will lead to much
improved limits on possible departures from the standard model. To this end,
I describe an analysis strategy to measure the $^{3}$\HeI\ flux emitted from nearby
low metallicity \HII\ regions. The proposed technique can be attempted with the next
generation of large telescopes, and will be easier to realize in metal-poor
\HII\ regions with quiescent kinematics.
\end{abstract}


\section{Introduction}

The standard model of cosmology and particle physics provides
our current best physical
description of the Universe. Two of the most remarkable
predictions of this model, now experimentally confirmed,
include the production of the lightest chemical elements
during the first minutes after the Big Bang (known as
Big Bang Nucleosynthesis, or BBN; \citealt{AlpBetGam48}),
and the existence of a relic cosmic microwave background
(CMB) radiation \citep{PenWil65}.

Seconds after the Big Bang, ordinary matter consisted mostly
of free neutrons and protons. As the Universe
expanded and cooled, some neutrons and protons combined
to form deuterium (D); soon after, BBN entered full production. Nearly
all of the deuterium nuclei were fused to form the more tightly bound
helium-4 ($^4$He) nuclide. Other light nuclides were also produced,
but in much lower abundance, including the lighter isotope of
helium ($^3$He) and a very small amount of lithium-7 ($^7$Li).
After $\sim\!\!20$ minutes, the nuclear reactions were quenched
by the declining temperature and density of the Universe, and
BBN was complete (for a review, see \citealt{Ste07,Cyb15}).

The relative abundances of the primordial nuclides produced
during BBN are sensitive to the universal expansion rate, and
the cosmic density of ordinary matter (i.e. the cosmic baryon
density, \ob). The expansion rate is set by the
total energy density of CMB photons and relativistic particles,
which for the standard model includes electrons, positrons and
three flavors of neutrino. Typically, the various contributions to the
expansion rate are collected and parameterized by an ``effective
number of neutrino species'', \neff, where the standard
model value corresponds to \neff~$=3.046$ \citep{Man05}.
Deviations from this value could indicate new physics not
presently captured by the standard model.

The dependence of each primordial element ratio on
\ob\ and \neff\ can be determined from detailed numerical
calculations of BBN. Historically, each primordial nuclide
is compared relative to hydrogen.
For example, the primordial deuterium abundance (D/H) has
commonly been used to estimate the cosmic baryon density
\citep{WagFowHoy67}, while the mass fraction of all baryons
consisting of $^{4}$He (\yp) depends strongly on \neff\
and is relatively insensitive to \ob\ \citep{SteSchGun77}.

Similar measurements of \ob\ and \neff\ are also available through
a careful analysis of the CMB temperature fluctuations, recently
recorded in exquisite detail by the \citet{Efs15}. Their analysis
provides \obhh(CMB)~$=0.02229^{+0.00039}_{-0.00040}$ and
\neff(CMB)~$=3.04\pm0.36$ (95 per cent confidence limits), where
$h$ is the Hubble constant in units of $100~{\rm km~s}^{-1}~{\rm Mpc}^{-1}$.
Recent measurements of D/H \citep{PetCoo12,Coo14} provide an independent
bound on \obhh(BBN) that is in good agreement with the values determined
from the CMB. However, the latest measurements of \yp\
suggest tentative, but inconclusive evidence favoring \neff(BBN)~$>$~\neff(CMB).
Specifically, the standard model value of the primordial $^4$He mass fraction,
inferred from the CMB, is \yp~$=0.24668\pm0.00007$ (68 per cent confidence
limits; \citealt{Efs15}). A recent survey conducted by \citet{IzoThuGus14} found
\yp~$=0.2551\pm0.0022$, which constitutes a $3.8\sigma$ deviation from the
standard model expectation. Using a subset of
the \citet{IzoThuGus14} survey data, \citet{AveOliSki15} derived a value
\yp~$=0.2449\pm0.0040$, which is more consistent with the standard model.
Thus, the different values of \yp\ derived by these authors might be due to
systematic differences in the analysis strategies adopted.

At present, it is unclear if additional, unaccounted for systematic uncertainties
are biasing the best measurements of \yp, and therefore masquerading
as non-standard physics (see e.g. Figure~8 from \citealt{Ste12}).
To alleviate this concern, a new, sensitive and reliable
probe is required to complement the measurement of \yp\ in the search for
possible departures from standard BBN. Until now, the power of combining
measurements of the primordial D/H and \heir\ ratios has not been fully
appreciated.

In this letter, I investigate the sensitivity and observational prospects
for uncovering new physics beyond the standard model using the D/H
and \heir\ abundance ratios set by BBN. In Section~\ref{sec:bbniso}, I derive the
dependence of these ratios on \obhh\ and \neff,
and summarize the current best observational determinations.
In Section~\ref{sec:future}, I discuss the importance of obtaining new
experimental values of several important reaction cross sections. I also
discuss the future potential of measuring the He isotope ratio in
near-primordial environments, before highlighting the main conclusions
of this work in Section~\ref{sec:conc}.

\begin{table}
\centering
\begin{minipage}[c]{0.5\textwidth}
    \caption{\textsc{H isotope ratio coefficients}}
    \begin{tabular}{lcccc}
    \hline
    \hline
    $m$ & $n$ & & & \\\cmidrule{2-5}
            & $0$ & $1$ & $2$ & $3$ \\
    \hline
    $0$& $29.428$ & $3.7763$ & $-0.18882$ & $0.045346$ \\
    $1$& $-3490.0$ & $-437.76$ & $38.547$ & $-8.7506$ \\
    $2$& $1.8127\times10^5$ & $21897.0$ & $-2849.5$ & $624.51$ \\
    $3$& $-4.4923\times10^6$ & $-5.1455\times10^5$ & $90285.0$ & $-19402.0$ \\
    $4$& $4.3247\times10^7$ & $4.6528\times10^6$ & $-1.0411\times10^6$ & $2.2120\times10^5$ \\
    \hline
    \end{tabular}
    \label{tab:hcoeff}
\end{minipage}
\end{table}


\section{The BBN Isotopes}
\label{sec:bbniso}

The relationships between \obhh, \neff, and the primordial abundances of D/H and \heir\
are derived from calculations \citep{Ioc09} that use the PArthENoPE code \citep{Pis08}.
These results are in good agreement with recent calculations \citep{Cyb15}
using the latest determinations of the neutron lifetime \citep{Oli15} and the
reaction rate cross-sections \citep{Xu13}. The primordial hydrogen and helium
isotope number abundance ratios are given by:
\begin{equation}\label{eqn:polyiso}
y_{\rm H, He} = \sum_{n}\sum_{m}\,a_{nm}\,\omega_{B}^{n}\,\Delta N_{\rm eff}^{m}
\end{equation}
where $y_{\rm H} = 10^{5}~\times$~D/H,
$y_{\rm He} = 10^{4}~\times$~\heir,
$\omega_{\rm B}=\Omega_{\rm B,0}\,h^{2}$, and $\Delta N_{\rm eff}~=~N_{\rm eff} - 3.046$.
The coefficients $a_{nm}$ are provided in Table~\ref{tab:hcoeff} and Table~\ref{tab:hecoeff}.
I adopt a conservative $5\%$ standard error in $y_{\rm H}$ \citep{Cyb15} and a
$3\%$ standard error in $y_{\rm He}$ \citep{Ste07} due to uncertainties in the nuclear
reaction rates relevant for the BBN calculations. As discussed further below, this
uncertainty dominates the current error budget on \obhh(BBN) and \neff(BBN).

The D/H number abundance ratio can be determined
in near-primordial environments to high precision. The employed technique \citep{Ada76}
requires a rare, chance alignment between a near-pristine ``cloud'' of neutral gas and
a bright, usually unrelated, background source (typically a quasar). The foreground gas
cloud imprints the Lyman series absorption lines of neutral D and H on the light from
the unrelated background quasar, allowing the relative column density of \DI\ and \HI\
to be measured. In this work, I adopt the latest determination
D/H~$=(2.53\pm0.04)\times10^{-5}$ by \citet{Coo14},
derived from the highest precision measurements
currently available.

\begin{table}
\centering
\begin{minipage}[c]{0.5\textwidth}
    \caption{\textsc{He isotope ratio coefficients}}
    \begin{tabular}{lcccc}
    \hline
    \hline
    $m$ & $n$ & & & \\\cmidrule{2-5}
            & $0$ & $1$ & $2$ & $3$ \\
    \hline
    $0$& $50.278$ & $-2.6670$ & $0.11296$ & $0.11005$ \\
    $1$& $-4705.2$ & $519.14$ & $-0.45387$ & $-30.061$ \\
    $2$& $2.3843\times10^{5}$ & $-41405.0$ & $-221.73$ & $2564.1$ \\
    $3$& $-5.9383\times10^{6}$ & $1.4212\times10^{6}$ & $5602.0$ & $-90887.0$ \\
    $4$& $5.7966\times10^{7}$ & $-1.7735\times10^{7}$ & $7849.5$ & $1.1535\times10^{6}$ \\
    \hline
    \end{tabular}
    \label{tab:hecoeff}
\end{minipage}
\end{table}

The helium isotope ratio (\heir), on the other hand, has received
relatively little attention from both the theoretical and observational
BBN community. The current best estimate of the
He isotope ratio is measured from the so-called
`quintessence' phase (He-Q) from solar system meteorite
samples \citep{LewSriAnd75}. The isotope ratios of the noble gases that are
incorporated into `phase Q' are believed to represent the
values that preceded the formation of the solar system,
when the Universe was less than 67 per cent of its
current age (i.e. less than $\sim9$~billion years after
the Big Bang). Although He-Q may not represent the
primordial \heir\ (see Section~\ref{sec:pristine}), this
estimate provides an illustrative value that can be
used in the present work.

\begin{figure*}
  \centering
 {\includegraphics[angle=0,width=180mm]{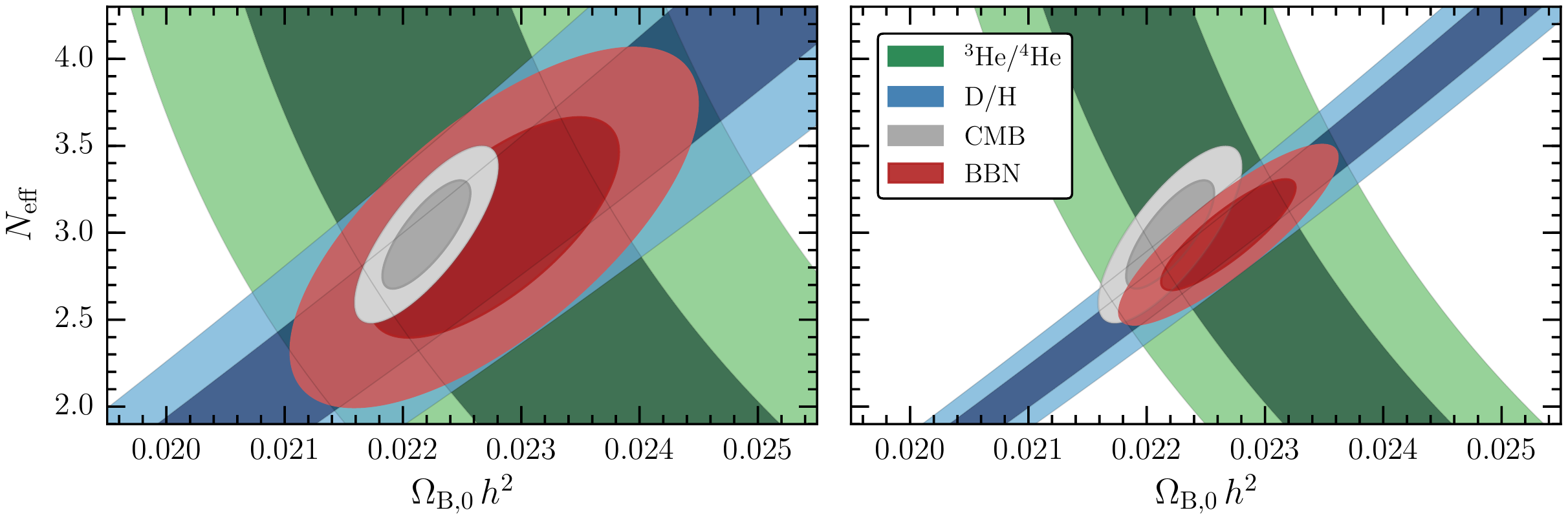}}\\
  \caption{
Combined confidence contours of the baryon density and the effective
number of neutrino species. The D/H (blue) and \heir\ (green)
contours are displayed, where dark and light shades represent the
68 and 95 per cent confidence contours respectively. Note that the
measured \heir\ isotope ratio adopted here is derived from solar
system meteorite samples, and is intended to be illustrative; this
determination of \heir\ may not reflect the primordial value. The gray contours
illustrate the results from the \textit{Planck} satellite observations of the
CMB temperature fluctuations \citep{Efs15}. The red contours show the
combined D/H and \heir\ (BBN only) confidence regions. The BBN contours
in the left panel use a $5\%$ and $3\%$ standard error
respectively for the BBN calculations, due to uncertainties in the nuclear
reaction rates. The right panel illustrates the same
contours, now assuming a $1\%$ uncertainty on the nuclear reaction rates,
and the \textit{same} observational measures. Note that the BBN contours in
the right panel are comparable in size to the latest CMB results.
  }
  \label{fig:contours}
\end{figure*}

Extracting the noble gases contained in phase Q requires
a two stage process \citep{BusBauWie00}.
In the first step, a meteorite is exposed
to hydrochloric and hydrofluoric acid (i.e. demineralization),
which leaves a resistant carbonaceous residue. This residue
is subsequently exposed to nitric acid, which oxidizes the
(presently unknown) carrier `Q', thereby releasing the noble
gases, including both He isotopes. The ideal meteorites for
determining the phase Q isotopic abundances are those that
have the lowest cosmic ray exposure age, since the isotope
ratio will be less affected by cosmogenic He production.
The best current determination of the He-Q isotope ratio
comes from the Isna meteorite \citep{BusBauWie00,BusBauWie01},
which has a remarkably short cosmic ray exposure age,
${\rm T}_{21} =150\,000~{\rm yrs}$ \citep{SchSch00}. The Isna
He-Q isotope ratio is $^{3}{\rm He}/^{4}{\rm He} = (1.23\pm0.02)\times10^{-4}$.

The BBN contours for the above observations and calculations are
presented in the left panel of Fig.~\ref{fig:contours}. This figure
highlights some of the benefits of using the He isotope ratio to
test the standard model; the D/H and \heir\ abundance ratios
offer almost orthogonal bounds on \obhh\ and \neff.
The determination of \neff\ therefore depends almost equally
on D/H and \heir, unlike the combination of D/H+\yp, where
\yp\ drives the determination of \neff.
Moreover, D/H, \yp, and \heir\ all exhibiting a very different
dependence on \obhh\ and \neff, and their combined measurement
will provide a highly complementary approach to identify
physics beyond the standard model.

Assuming that the pre-solar \heir\ ratio reflects the primordial value,
and adopting a conservative uncertainty in the BBN calculations
($5\%$ for $y_{\rm H}$, $3\%$ for $y_{\rm He}$), the following
limits are placed on the baryon density and the effective number
of neutrino species:
\obhh(BBN)~$=~0.0227^{+0.0016}_{-0.0013}$ and
\neff(BBN)~$=~3.01^{+0.95}_{-0.76}$ (95 per cent confidence limits).
These results agree remarkably well with the standard model of
cosmology and particle physics, and the \textit{Planck} CMB
analysis (gray contours in Fig.~\ref{fig:contours}).


\section{Future Prospects}
\label{sec:future}

\subsection{Improving the BBN Reaction Rates}

By improving a few key BBN nuclear reaction rates (discussed further below),
significantly tighter bounds on \obhh(BBN) and \neff(BBN) will be possible
using observational measures of comparable precision. For example, assuming
the same central values for the BBN reaction rates, a 1 per cent
standard error in these rates would reduce the uncertainty on
\neff(BBN) by a factor of 2. The BBN contours for this level of uncertainty
in the reaction rates are illustrated in the right panel of Fig.~\ref{fig:contours},
and are competitive with that of the \textit{Planck} CMB experiment.

In order for this level of precision to be realized, several
BBN reaction rates must be revisited and measured using
modern facilities. The three most uncertain
reaction rates that determine the D/H abundance are d(p,$\gamma)^{3}$He,
d(d,n)$^{3}$He, and d(d,p)$^{3}$H \citep{Fio98,NolBur00,Cyb04,Ser04}.
For the He isotope ratio, the most important reactions are d(p,$\gamma)^{3}$He
and $^{3}$He(d,p)$^{4}$He. New experimental data for the crucial
d(p,$\gamma)^{3}$He reaction, which contributes the dominant
uncertainty for both the D/H and \heir\ abundance ratios, are currently being acquired by the
Laboratory for Underground Nuclear Astrophysics (LUNA; \citealt{Bro10,DiV14}).
New data for the remaining reaction rates are now needed.


\subsection{Measuring \heir\ in near-pristine environments}
\label{sec:pristine}

\begin{figure*}
  \centering
 {\includegraphics[angle=0,width=180mm]{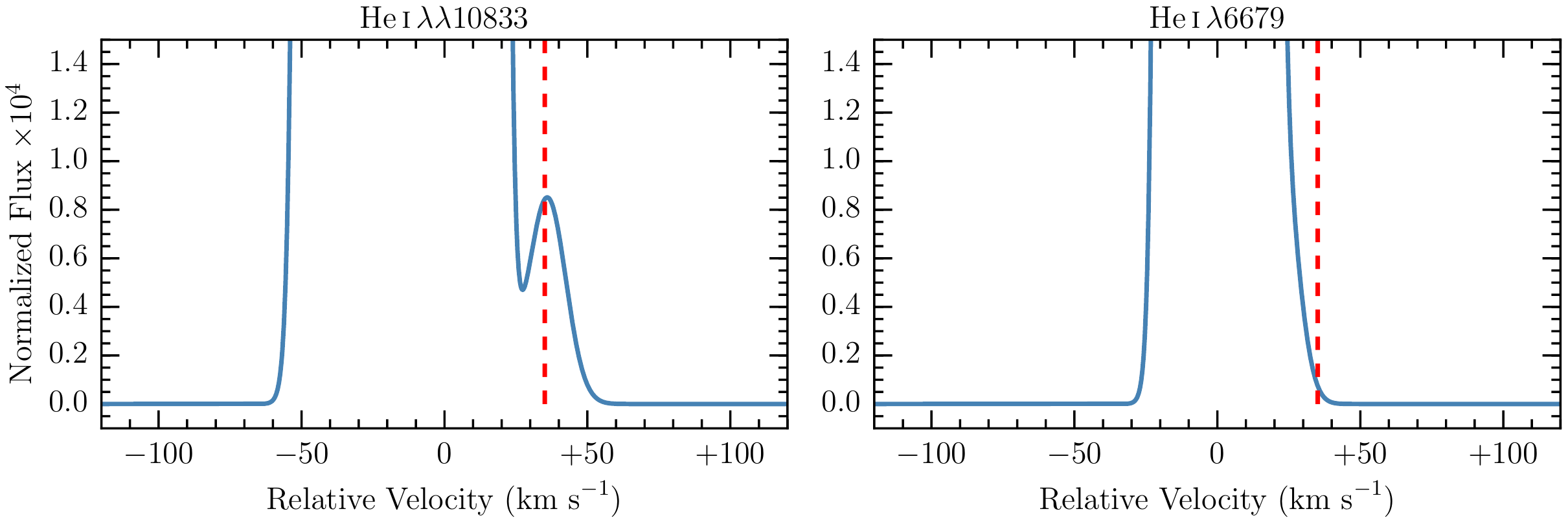}}\\
  \caption{
Synthetic line profiles for the \HeI\ triplet ($\lambda\lambda10833$; left panel)
and singlet ($\lambda6679$; right panel) emission.
Both profiles are normalized to the peak flux of the profile.
The vertical red dashed lines at a velocity of
$+35.2~{\rm km~s}^{-1}$ in both panels indicate the He
isotope shift of the $\lambda\lambda10833$ line. Note that the
He isotope shift of the $\lambda6679$ line is $+22.5~{\rm km~s}^{-1}$.
Therefore, the detection of an emission feature at $+35.2~{\rm km~s}^{-1}$
in the $\lambda\lambda10833$ profile, together with the absence of a feature
at the same velocity in the $\lambda6679$ profile, can be used to confirm the
identification as $^{3}$\HeI\ emission.
  }
  \label{fig:profiles}
\end{figure*}

The Isna He-Q measurement currently provides our best indication of
the primordial He isotope ratio. However, it remains unclear if He-Q
is a true reflection of the primordial BBN value.
A measurement of the He isotope ratio in a near-pristine
environment, where the post-BBN chemical evolution of $^{3}$He is expected
to be negligible, will provide a more reliable determination of the
truly primordial $^{3}$He\,/\,$^{4}$He ratio.
In principle, such a measurement should be possible in known
metal-poor \HII\ regions by comparing the profiles of two \HeI\
emission lines, since the isotope shift is different for all
\HeI\ transitions (a relative shift of up to $40~{\rm km~s}^{-1}$ for the
optical and near-infrared emission lines; \citealt{MorWuDra06}). For
example, the $^{3}$He ${\rm 3D}\to{\rm 2P}$ singlet transition
(\HeI\,$\lambda6679$) has an isotope shift of
$\simeq+22.5~{\rm km~s}^{-1}$, and the
$^{3}$He ${\rm 2P}\to{\rm 2S}$ triplet transition
(\HeI\,$\lambda\lambda10833$) has an
isotope shift of $\simeq+35.2~{\rm km~s}^{-1}$.

The above combination of emission lines is perhaps the best suited to obtain
a direct measurement of the $^{3}$\HeI\ flux from a metal-poor \HII\ region.
Assuming a flux ratio of ${\cal F}(^3{\rm He}/^{4}{\rm He})=10^{-4}$, the
centroid of the $^{3}$\HeI\ line would need to be ideally $\gtrsim5\sigma$ from the centroid
of the $^{4}$\HeI\ line to ensure that the $^{3}$\HeI\ emission dominates the signal.
For the $1.0833~\mu$m \HeI\ line, this $\gtrsim5\sigma$ shift corresponds to a $^{4}$\HeI\ profile
with a full width at half maximum, FWHM~$\lesssim18~{\rm km~s}^{-1}$.
For comparison, thermal broadening of the $^{4}$He line at a temperature
of 15,000~K, which is typical of \HII\ regions \citep{IzoThuGus14,AveOliSki15},
would produce a profile with FWHM~$\simeq13~{\rm km~s}^{-1}$.

To detect the $^{3}$\HeI\ feature at $5\sigma$ confidence,
the signal-to-noise (S/N) ratio of the $^{4}$\HeI\ feature must exceed
$5\times\sqrt{{\cal F}(^4{\rm He}/^{3}{\rm He})}$, or S/N~$\gtrsim500$.
This S/N requirement would need to be increased if the stellar continuum of the target
galaxy is bright, or if the red wing of the $^4$\HeI\ emission line
extends over the $^3$\HeI\ emission line.
In any case, the primary requirement is that the $^4$\HeI\ profile is smooth
and narrow. To confirm that an observed feature is due to $^{3}$\HeI\
emission, and not a coincident weak $^{4}$\HeI\ emission feature,
another \HeI\ emission line with a different isotope shift (e.g. \HeI\,$\lambda6679$)
can be inspected. An example comparison of two \HeI\ emission lines
is presented in Fig.~\ref{fig:profiles} for a Gaussian profile
with FWHM~$=13~{\rm km~s}^{-1}$.

Measuring the $^{3}$\HeI\ flux from a nearby metal-poor \HII\ region
might be possible with the next generation of  30+\,m telescopes.
The current instrument concepts of the HIRES and HROS echelle
spectrographs, to be mounted on the European Extremely Large
Telescope and the Thirty Meter Telescope respectively, are
well-designed for this experiment. The wavelength range covered
by these instrument concepts includes the best available \HeI\
emission lines with a single instrument setup.
Moreover, the large telescope aperture greatly facilitates the final S/N
that can be achieved. However, to realize this measurement,
emphasis should now be placed on discovering suitable low
metallicity \HII\ regions with quiescent kinematics.


\section{Summary and Conclusions}
\label{sec:conc}

The primordial abundance of helium-3 ($^{3}$He/H), is generally considered to
provide very little constraining power on the physics of BBN,
since it is less sensitive to the baryon density than the primordial deuterium abundance, is
less affected by the expansion rate than the $^{4}$He mass fraction, and a reliable estimate
of the primordial value has not been measured. This paper
highlights the possibility of combining the deuterium abundance and the helium
isotope ratio (\heir) to simultaneously estimate \obhh\ and \neff\ at the time of BBN.
The following conclusions are drawn:\\

\noindent ~~(i) The D/H and \heir\ isotope ratios set by BBN offer almost
orthogonal bounds on \obhh\ and \neff, unlike Y$_{\rm P}$ which
is essentially independent of \obhh. Therefore, by combining D/H
and \heir, evidence for physics beyond the standard model
is less influenced by a single measurement.

\smallskip

\noindent ~~(ii) If the pre-solar meteoritic value of \heir\ reflects the
primordial value, the following bounds can be placed on the baryon
density, \obhh(BBN)~$=0.0227^{+0.0016}_{-0.0013}$,
and the effective number of neutrino species,
\neff(BBN)~$=3.01^{+0.95}_{-0.76}$ with 95 per cent confidence,
assuming a conservative uncertainty for the measured
BBN reaction rates ($5\%$ for $y_{\rm H}$ and $3\%$ for $y_{\rm He}$).

\smallskip

\noindent ~~(iii) In order to achieve bounds on \obhh\ and \neff\ that are
competitive with the
latest results from the \textit{Planck} satellite, several important BBN
reaction rates must be redetermined, including d(p,$\gamma)^{3}$He,
d(d,n)$^{3}$He, d(d,p)$^{3}$H, and $^{3}$He(d,p)$^{4}$He.

\smallskip

\noindent ~~(iv) It is presently unclear to what degree the He-Q isotope
ratio is affected by post-BBN nucleosynthesis. It is therefore necessary
to obtain a measurement of the He isotope ratio in a
near-pristine environment. Nearby metal-poor \HII\ regions could be
the most promising systems to estimate the primordial $^{3}$He/$^{4}$He
ratio. The \HeI\ optical and near-infrared emission lines typically
seen in \HII\ regions exhibit a variety of isotope shifts (up to $\sim$40~km~s$^{-1}$),
allowing the $^{3}$He emission to be unambiguously identified.
With a view to this goal, I present a possible
strategy to measure the $^{3}$\HeI\ flux from
nearby metal-poor \HII\ regions.

\smallskip

Although such a delicate measurement will have to wait for the next
generation of $30+$\,m telescope facilities, it is critical that the relevant
BBN reaction rates are now measured with high precision. Once this
goal is achieved, the combined information provided by the primordial
D/H abundance and the \heir\ isotope ratio has the potential to deliver
a reliable probe of possible departures from the standard model of
physics during the early Universe.

\acknowledgments

I thank M.~Pettini and J.~X.~Prochaska for useful
discussions about the work described in this paper,
and for suggesting comments on an earlier draft.
I am grateful to the anonymous referee who provided
constructive comments that improved the presentation
and clarity of this work.
R.~J.~C. is currently supported by NASA through
Hubble Fellowship grant HST-HF-51338.001-A, awarded by the
Space Telescope Science Institute, which is operated by the
Association of Universities for Research in Astronomy, Inc.,
for NASA, under contract NAS5- 26555. The majority of the
analysis and figures presented in this paper were prepared using
publicly available \textsc{python} packages, including Astropy, NumPy,
SciPy, Cython, and Matplotlib \citep{ast13,vdWColVar11,Beh11,Hun07}.

\end{document}